\documentclass[%
 aip,
cp,  
 amsmath,amssymb,
 reprint,%
]{revtex4-2}
\usepackage{graphicx}
\usepackage{dcolumn}
\usepackage{bm}
\usepackage[center]{subfigure}
\usepackage[utf8]{inputenc}
\usepackage[T1]{fontenc}
\usepackage{mathptmx} 

\begin{document}

\title{Studies for New Experiments at the CERN M2 Beamline within ``Physics Beyond Colliders'':\\AMBER/COMPASS++, NA64$\mu$, MuonE}

\author{Johannes Bernhard}
 \email[Corresponding author: ]{johannes.bernhard@cern.ch}
 \affiliation{CERN}
\author{Dipanwita Banerjee}
\affiliation{CERN}
\affiliation{University of Illinois at Urbana-Champaign}
\author{Eva Montbarbon}
\author{Markus Brugger}
\author{Nikolaos Charitonidis}
\author{Serhii Cholak}
\affiliation{CERN}
\author{Gian Luigi D'Alessandro}
\affiliation{CERN}
\affiliation{Royal Holloway, University of London}
\author{Lau Gatignon}
\author{Alexander Gerbershagen}
\author{Bastien Rae}
\author{Marcel Rosenthal}
\author{Maarten van Dijk}
\affiliation{CERN}
\author{Benjamin Moritz Veit}
\affiliation{CERN}
\affiliation{Institut f\"ur Kernphysik, Johannes Gutenberg-Universit\"at Mainz}

\date{\today}

\begin{abstract}
The ``Physics Beyond Colliders (PBC)'' study explores fundamental physics opportunities at the CERN accelerator complex complementary to collider experiments. Three new collaborations aim to exploit the M2 beamline in the North Area with existing high-intensity muon and hadron beams, but also aspire to go beyond the current M2 capabilities with a RF-separated, high-intensity hadron beam, under study. The AMBER/COMPASS++ collaboration proposes an ambitious program with a measurement of the proton radius with muon beams, as well as QCD-related studies from pion PDFs / Drell-Yan to cross section measurements for dark sector searches. Assuming feasibility of the RF-separated beam, the spectrum of strange mesons would enter a high precision era while kaon PDFs as well as nucleon TMDs would be accessible via Drell-Yan reactions. The NA64$\mu$ collaboration proposes to search for dark sector mediators such as a dark scalar $A'$ or a hypothetical $Z_\mu$ using the M2 muon beam and complementing their on-going $A'$ searches with electron beams. The MuonE collaboration intends to assess the hadronic component of the vacuum polarization via elastic $\mu-e$ scattering, the dominant uncertainty in the determination of $g_\mu-2$. An overview of the three new experimental programs will be presented together with implications for the M2 beamline and the experimental area EHN2, based on the studies of the PBC ``Conventional Beams'' Working Group.
\end{abstract}
\maketitle

\section{\label{sec:PBC}The Physics Beyond Colliders Study}
Physics Beyond Colliders is an exploratory study aiming at exploiting the full scientific potential of CERN's accelerator complex and its scientific infrastructure through projects complementary to the LHC, HL-LHC and other possible future colliders. The proposed projects shall target fundamental physics questions that are similar in spirit to those addressed by high-energy colliders, but that require different types of beams and experiments. The study was initiated in 2016 by the CERN director-general Fabiola Gianotti and is coordinated by J\"org J\"ackel, Mike Lamont and Claude Vallee. During the kick-off workshop in September 2016, numerous areas of interest were identified, ranging from QCD to Beyond the Standard Model physics. First conclusive results and summary presentations were presented during the last PBC annual meeting in January 2019 and were summarised in a full report\cite{Jaeckel:2651120}. Most proposals have been brought forward to the European Strategy for Particle Physics\footnote{Input to the European Particle Physics Strategy Update, https://indico.cern.ch/event/765096/contributions/, accessed on 15.08.2019}, which was discussed recently in an open symposium organised by the CERN Council in Granada\footnote{CERN Council Open Symposium on the European Particle Physics Strategy Update,  https://cafpe.ugr.es/eppsu2019/, accessed on 15.08.2019}. A physics briefing book is currently being prepared, summarising the proposals that will be discussed during the strategy drafting session beginning of 2020 in Bad Honnef and then finally decided upon by the CERN council in May 2021.
\vspace{-1em}
\subsubsection{Structure and Working Groups}
\begin{figure}[tbp]
    \centering
    \vspace{-1em}
    \begin{subfigure}{}
    \includegraphics[width=0.45\linewidth]{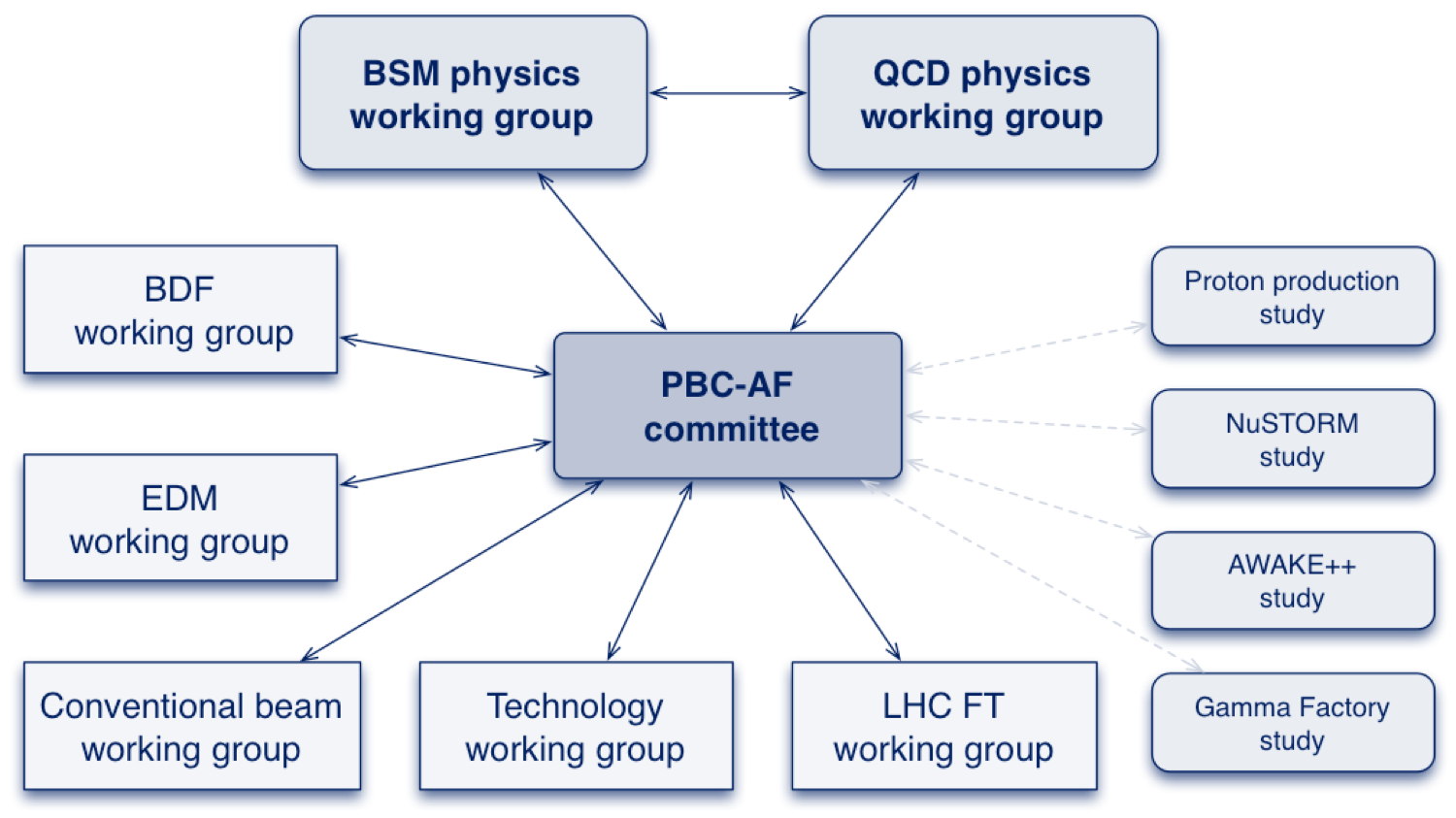}
    \end{subfigure}
    \begin{subfigure}{}
    \includegraphics[width=0.45\linewidth]{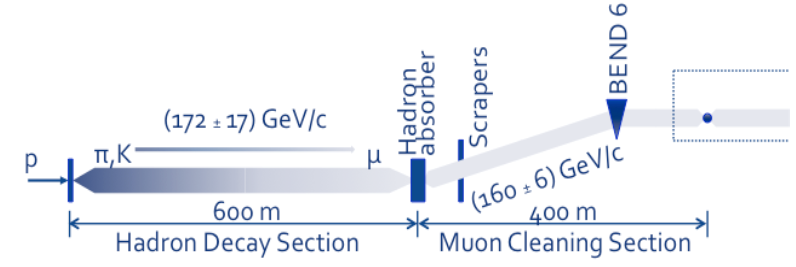}
    \end{subfigure}
    \caption{Left: Structure and Working Groups of the ``Physics Beyond Colliders'' Study. Right: Schematic layout of the M2 beamline. The specific momenta for $160~\mathrm{GeV/}c$ muon beams are given as an example.}
    \label{fig:M2-layout}
    \vspace{-1em}
 \end{figure}

\vspace{-0.5em}
Several working groups pursue feasibility and case studies for the proposed projects, see left panel of Fig.~\ref{fig:M2-layout}. The physics groups have been established according to the main topics ``Quantum Chromodynamics (QCD WG)''\cite{Diehl:2652442} and ``Physics beyond the Standard Model (BSM WG)''\cite{Beacham:2652223}, while several working groups have been formed under steering of the ``Accelerator and Facilities'' committee. One of these groups is the ``Conventional Beams'' working group, which performs technical studies for the proposed projects in CERN's North and East Area, i.e. for mostly fixed-target experiments. A dedicated subgroup has been formed for the three proposals concerning the M2 beamline, serving the EHN2 experimental area that currently hosts the COMPASS experiment. We will discuss briefly physics motivations, technical aspects and the relevant studies for these proposals within the EHN2 subgroup in the following.
 
\section{The M2 beamline}
\vspace{-0.5em}The M2 beamline~\cite{Doble:250676} at the North Area delivers high-intensity and high-energy
muon and hadron beams towards the experimental area EHN2 as well as low-intensity electron beams for calibration purposes. The schematic layout of M2 is
shown in Fig.~\ref{fig:M2-layout}, right panel.
The M2 beamline transports secondary particles from the T6 target over a total distance of about 1130~m to EHN2. The front-end of the beamline consists of six high-gradient quadrupoles with a high acceptance, mainly driven by the required intensities of secondary pions for the muon beam mode. Afterwards, the momentum selection is performed in the horizontal plane with a maximum momentum bite of $\pm 10\% ~\Delta p/p$. 
A long FODO section of approximately 600~m length follows in order to allow for the decay of secondary particles into muons. Depending on the operational mode (hadron/muon beams), a set of nine motorised Beryllium absorbers with a maximum total length of 9.9~m can be inserted in a series of vertical dipole magnets to absorb remaining hadrons. 
The remaining hadron contamination is of the order of about $10^{-6}$ to $10^{-5}$. Equipped with magnetic collimators, another 400~m long FODO section performs the final muon momentum selection in combination with subsequent dipoles and is used to clean the beam from muon halo, i.e. scattered muons that are nevertheless transported throughout return fields in magnet yokes. In addition, scintillator hodoscopes for momentum measurements (``BMS'') and beam particle identification via two differential Cherenkov detectors (CEDARs) are available in this section of the beamline. The beam is then focused onto the experiment location in EHN2. Different modes of operation are summarised in Table~\ref{tab:M2-operation-modes}, as employed in the last years for the COMPASS experiment.
 
 \begin{table}[htbp]
    \centering
    \caption{Summary of the available operation modes of the M2 beamline. For tertiary beams, the secondary beam momentum is listed, as well.}
       \begin{tabular}{lllllll}
       \hline \hline
       \textbf{Beam}      & \textbf{Polarity and Momentum $qp/e$}   & \textbf{Max. Flux}                & \textbf{Typical}             & \textbf{Typical RMS}    & \textbf{Polarisation} & \textbf{Absorber}                \\
       \textbf{Mode}      & \textbf{[$\mathrm{GeV}/c$]} & \textbf{[$\mathrm{ppp}/4.8~\mathrm{s}$]} & \textbf{[$\Delta p/p (\%)$]} & \textbf{spot at target} &                       & \textbf{[$9.9~\mathrm{m~Be}$]} \\
       \hline
       Muons              & $+208/+190$          & $\sim 10^8$                       & $3\%$                        & $8\times8$~mm$^2$         & $80\%$                & IN                               \\
                          & $+172/+160$          & $2.5\cdot10^8$                    &                              &                         &                       &                                  \\
       \hline
                          & $+190$              & $10^8$ (RP)                       &                              &                         &                       &                                  \\
       Hadrons            & $-190$              & $4\cdot10^8$ (with                & -                            & $5\times5$~mm$^2$         & -                     & OUT                              \\
                          & Max. $280$          & dedicated dump)                   &                              &                         &                       &                                  \\
       \hline
       Electrons          & $-10$ to $-40$      & $< 2\cdot10^4$                    & -                            & $> 10\times10$~mm$^2$     & -                     & OUT                              \\
       \hline \hline
    \end{tabular}
    \label{tab:M2-operation-modes}
 \end{table}
 
\vspace{-0.5em}
\section{NA64$\mu$}
\vspace{-0.5em}
The NA64 collaboration currently searches for light dark matter candidates with the help of tertiary electron beams at the H4 beamline in the EHN1 experimental area at CERN\cite{Gninenko:1634093}. These searches are performed in an active beam dump experiment (NA64e) employing the missing energy technique, and concentrate around the mass range of about 1~MeV$/c^2$ to 1~GeV$/c^2$. While the NA64e experiment was recently approved to continue\cite{Gninenko:2300189} in 2021, the collaboration proposes to extend the programme towards muon beams\cite{Gninenko:2653581}. In addition to extending the accessible mass range for the light dark matter searches, exploiting muon beams would enable the search for a possible light gauge boson, which would couple pre-dominantly to muons and/or taus. Such a light boson could be either a scalar (S$_\mu$) or vector (Z$_\mu$) mediator and would be a viable explanation\cite{Gninenko:2001119} for the long-standing 3.6$\sigma$ discrepancy between the measured\cite{PhysRevLett.89.101804, PhysRevLett.92.161802, PhysRevD.73.072003} and theoretically calculated\cite{Jegerlehner:20091, PhysRevD.97.114025} values of the anomalous magnetic moment of the muon, especially as an explanation via a dark photon $A'$ could be already ruled out in the invisible decay channel by combined data from NA64e\cite{PhysRevD.97.072002}, BABAR\cite{PhysRevLett.119.131804}, and NA62\cite{CortinaGil2019}.

\begin{figure}[tbp]
    \centering
    \vspace{-1em}
    \includegraphics[width=0.8\linewidth]{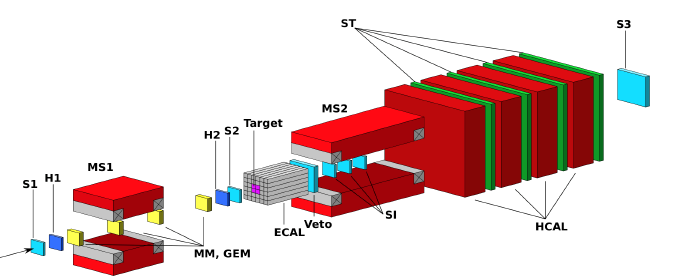}
    \caption{Layout of the proposed first phase for the NA64$\mu$ experiment in the upstream part of the M2 beamline.}
    \label{fig:NA64setup}
     \vspace{-1em}
 \end{figure}

\begin{figure}[tbp]
    \centering
    \vspace{-1em}
    \includegraphics[width=0.7\linewidth]{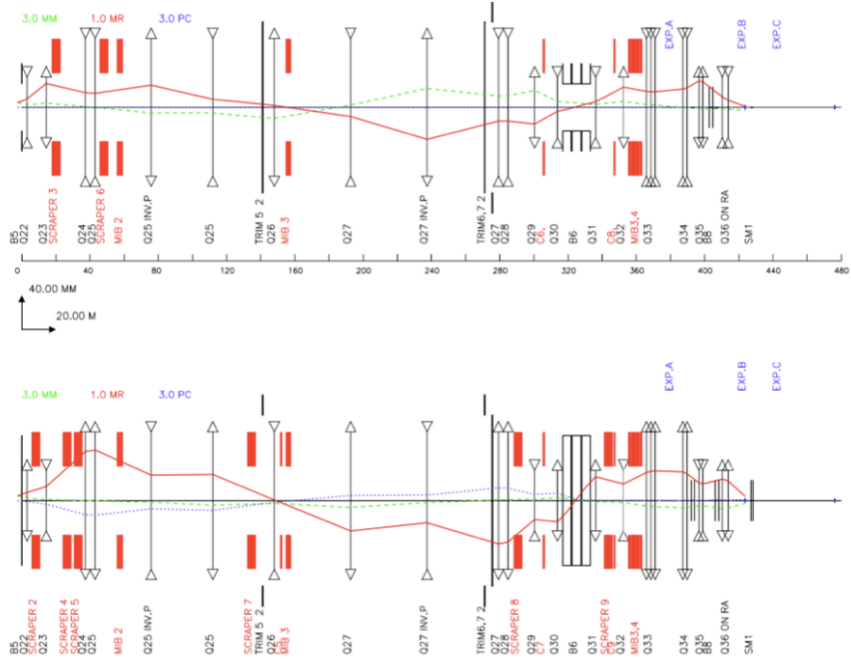}
    \caption{Tentative optics for the NA64$_\mu$ phase 1 location in the upstream part of the EHN2 area. The red line line corresponds to the sine-like ray, the green line represents the magnification term and the dotted-blue line is the dispersive term. Calculations were performed with {\tt TRANSPORT}\cite{Brown:133647}.}
    \label{fig:NA64optics}
    \vspace{-1em}
 \end{figure}

The proposed experiment is divided into two phases. The first phase will concentrate on the search for Z$_\mu$, which would be produced in a Bremsstrahlung process of the 160 GeV$/c$ muon beam along an active target. The set-up is depicted in Fig.~\ref{fig:NA64setup}. The beam momentum is first determined with the help of the existing COMPASS beam momentum station (BMS, not depicted), which surrounds the last strong, vertical dipole in the beamline. The momentum measurement is then crosschecked with another magnetic spectrometer (MS1), employing MircoMega and GEM detectors for tracking and scintillator counters for the trigger system. The active target is a shashlyk calorimeter module of 40 X$_0$ length, surrounded hermetically by additional shashlyk modules to detect lateral energy leaks from the target (ECAL). The outgoing momentum of the scattered muon is determined by another magnetic spectrometer (MS2), which is equipped with Veto scintillators for matching the target acceptance and large angle scattering outside of the acceptance of the hadronic calorimeter (HCAL), following directly MS2. In between the HCAL modules Straw detectors are available for tracking and vetoing of muons. The signature for Z$_\mu$ production would be the detection of a scattered muon with a detected energy loss of more than 50\% of the beam energy, accompanied by no other signals from HCAL and Vetos.\\
For implementation of the experiment in EHN2, two options were studied\cite{Gatignon:2650989, Banerjee:IPAC2019-THPGW063}. The first option is the installation of NA64$\mu$ downstream of the existing COMPASS experiment. This option was discarded, as it is not possible to refocus the beam at this location without interference with the COMPASS experimental programme. The COMPASS spectrometer represents also a non-negligible amount of material in the muon beam, which would lead to multiple scattering. Also, only about 15~m of space are available in the remaining part of the EHN2 area, which is incompatible with the required space, and costly infrastructure for the required additional magnets for MS1 and MS2 would have to be installed. Nevertheless, this space turned out to be very useful for first tests in 2018 that were done parasitically to the COMPASS data taking. The second option would be the installation directly at the exit of the M2 beamline tunnel, upstream of the COMPASS experiment. This is the preferred option as beamline and detector elements can be added and removed comparatively easily and enough space of about 40~m would be available thanks to the space reserved for the CEDAR beam instrumentation, which is not required for muon beam operation.\\
Beam optics were studied with the help of {\tt TRANSPORT}\cite{Brown:133647}, {\tt TURTLE}\cite{Brown:186178} and {\tt HALO}\cite{Iselin:186209} with the aim to have a parallel, small beam as required by the experiment. The tentative optics are shown in Fig.~\ref{fig:NA64optics}. We obtain a beam spot with a size of 20~mm ($\sigma_x$) $\times$ 20~mm ($\sigma_y$) and a divergence of 0.2~mrad ($\sigma_{x'}$) $\times$ 0.2~mrad ($\sigma_{y'}$) at the location of the active target. The 160~GeV$/c$ beam has a width of 6 GeV$/c$ ($\sigma_p$). In order to study the feasibility of a second experiment at a more downstream location, e.g. at the COMPASS target, a check of the beam contamination due to the NA64$\mu$ detector material in the beamline was performed with {\tt FLUKA}\cite{fluka1, fluka2} and {\tt GEANT4}\cite{geant4}. Only a negligible contamination by hadrons of less than $10^{-5}$ has been found. Nevertheless, running of a downstream experiment can be excluded due to the demanding beam requirements for beam spot and divergence of the proposals in question that would be degraded due to multiple scattering in the NA64$\mu$ detectors. In the second phase, the collaboration would continue with searches for milli-charged particles, $\mu - \tau$ conversion in flight and further searches for light dark matter mediators. The preferred location for this experiment would be inside the large SM2 spectrometer magnet that is currently being used by COMPASS. Studies are on-going\cite{Gatignon:2650989, Gninenko:2653581} while the programme is still evolving.

\vspace{-0.5em}
\section{MuonE}
\begin{figure}[tbp]
    \centering
    \vspace{-1em}
    \includegraphics[width=0.8\linewidth]{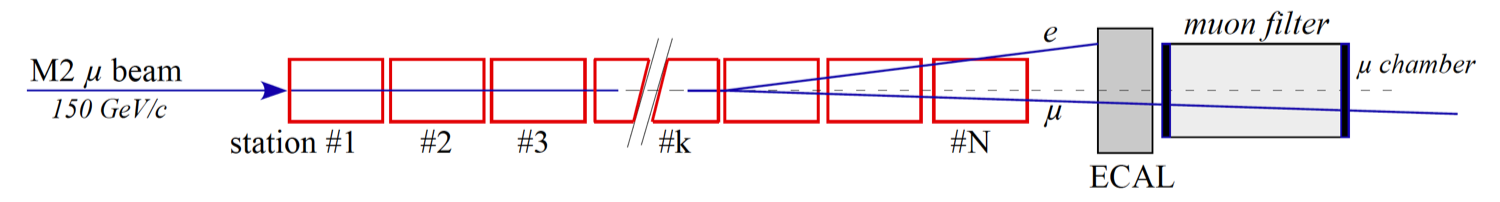}
    \caption{Intended set-up for the MuonE experiment in the upstream location of EHN2.}
    \label{fig:muonesetup}
    \vspace{-1em}
 \end{figure}

\vspace{-0.5em}
As already explained in the previous section, the deviation of measured and predicted value of $(g-2)_\mu$ is still puzzling. One important ingredient to the error budget of both theory and experiment is the knowledge of the hadronic vacuum polarisation. So far, the value of this QCD correction could be obtained reasonably precisely by analysis of e$^+$e$^-$ scattering data, see e.g. \cite{Jegerlehner:20091, Davier2017}, but is limited due to the treatment of emergent resonances in the s-channel. Despite on-going, long-term efforts of the lattice community to reach a high-precision calculation (see e.g. \cite{Gerardin:2018sin, Meyer:2018til}), it might take several years to reach the impressive precision of the upcoming expected results of the experiments at Fermilab\cite{Grange:2015fou} and J-PARC\cite{Abe:2019thb}.
The MuonE collaboration intends\cite{Abbiendi:2677471} as an alternative approach the measurement of the hadronic component of the running electromagnetic
coupling $\alpha(t)$ in a momentum transfer $(t)$ region that is  relevant for the calculation of the muon g-2 anomaly via a measurement of elastic muon-electron scattering with a statistical accuracy of about 0.3\%. The experiment would be able to cover about 87\% of the $t$ region, while the remaining 13\% could be determined using pQCD and time-like data and/or lattice QCD results.
The intended experimental set-up is depicted in Fig.~\ref{fig:muonesetup}. It comprises a series of thin targets with a low-Z material, such as Beryllium or Carbon, in order to minimise multiple scattering. The targets are interleaved with Silicon trackers originally developed for CMS upgrade II, which measure the tracks of the outgoing muon and electron from the elastic interaction. After the last target-detector station, an electromagnetic calorimeter for systematic checks and a muon filter are planned.
The calculated optics for the 150 GeV$/c$ muon beam are very similar to the ones depicted in Fig.~\ref{fig:NA64optics} due to the similar requirements. Main differences are due to the different lateral size of MuonE with respect to NA64$\mu$, which results in slightly different settings of the magnetic collimators in the M2 beamline. The larger acceptance of about $10 \times 10$~cm$^2$ allows for highest intensities, as the beam does not have to be collimated to reach a small size while keeping the beam parallel.
The major challenge of the measurement is the control of both theoretical and experimental systematics at a comparable level of accuracy. Due to the required overall experimental precision in the order of $10^{-5}$, the technical implementation of the experiment is demanding, for instance in terms of alignment and thermal stability.\\
A pilot run with two target-detector stations was requested for 2021 in order to check the systematic error budget in realistic conditions and demonstrate the feasibility of the several engineering challenges.

\vspace{-0.5em}
\section{AMBER/COMPASS++}
\begin{figure}[tbph]
    \centering
    \vspace{1em}
    \includegraphics[width=0.8\linewidth]{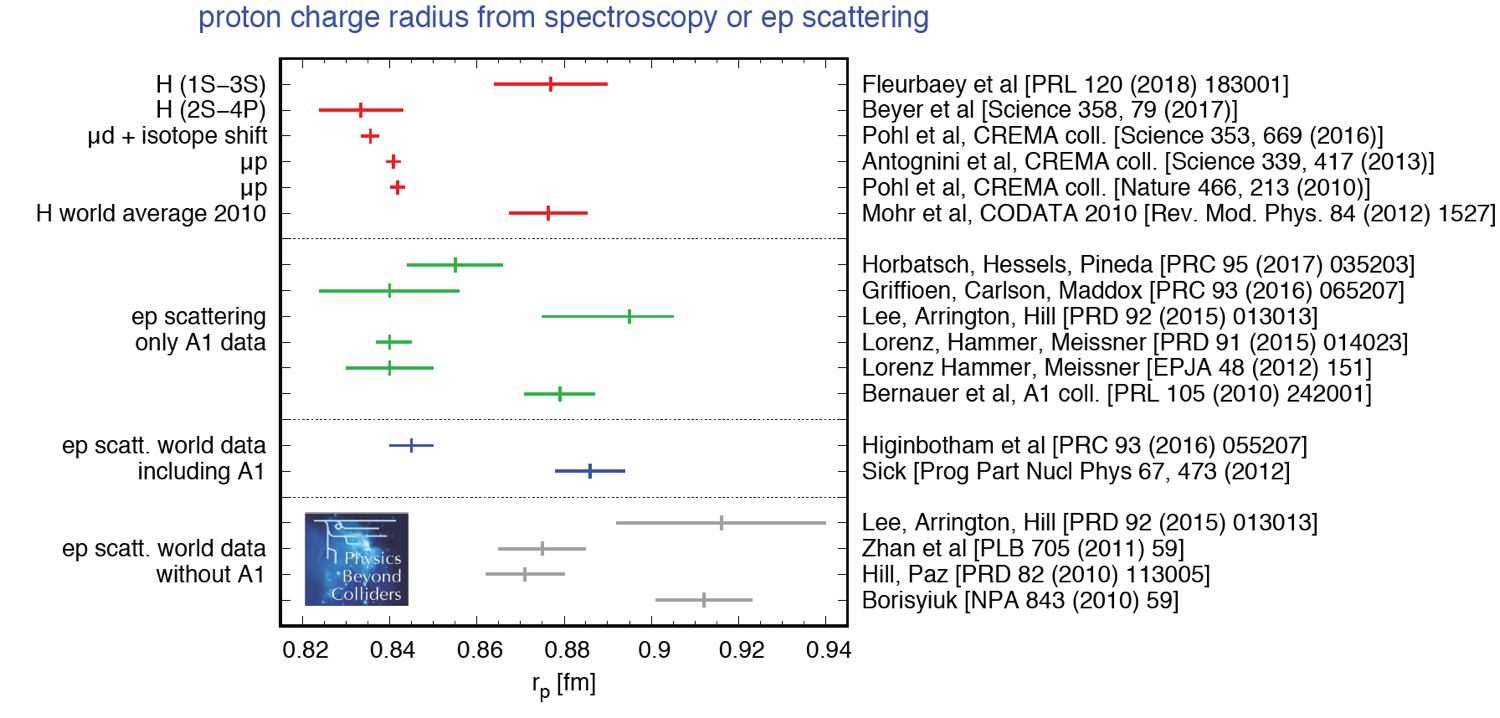}
    \caption{Overview on recent measurements of the proton radius as taken from \cite{Jaeckel:2651120}.}
    \label{fig:pradoverview}
    \vspace{-1em}
 \end{figure}

\vspace{-0.5em}
The AMBER/COMPASS++ collaboration proposes an ambitious programme for a new QCD facility at CERN\cite{Adams:2676885, Adams:2653603}, making use of existing COMPASS detector infrastructure together with several upgrades and new detectors. In the first phase, the collaboration focuses on a measurement of the proton charge radius with muon-proton elastic scattering, a measurement of proton-induced antiproton production cross sections for dark matter searches, and Drell-Yan and $J/\Psi$ production using the existing M2 hadron beam.\\
As a selected topic, we present here shortly the proton charge radius measurement, which aims to help resolving the so-called proton radius puzzle. The proton charge radius has been determined exploiting different methods, which yield conflicting results of spectroscopy vs. scattering measurements on the one hand and measurements with electrons vs. muons on the other hand. Figure~\ref{fig:pradoverview} summarises the currently available results, showing differences of up to 5 standard deviations.
The proposed measurement aims at a statistical accuracy of 0.01 fm or better and considerably smaller systematic uncertainty, the latter exploiting the fact that using muons, systematic effects and theoretical/radiative corrections are considerably smaller with respect to electron scattering. The experiment will comprise a time-projection chamber (TPC) as an active target, which is filled with pure hydrogen up to pressures of 20 bar. This allows to measure the recoil of the struck proton, which is used to determine $Q^2$ on an event-by-event basis. The envisaged energy for the muon beam is 100 GeV$/c$. As the location of the high-pressure TPC is the same as the currently used COMPASS target position, the existing muon beam optics are already fulfilling experimental requirements. Depending on evolving details of the TPC and beam telescope design, divergence and spot size of the beam might have to be slightly adapted. For 2021, test measurements with the newly constructed TPC are planned in the upstream location of the CEDARs, which were already removed for the 2021 data taking of COMPASS. An additional aim for the tests is the check of a new triggerless readout system. 

\vspace{-0.5em}
\subsubsection{RF-separated Beams}
\vspace{-0.5em}

\begin{figure}[tbph]
    \centering
    \vspace{-1em}
    \includegraphics[width=0.8\linewidth]{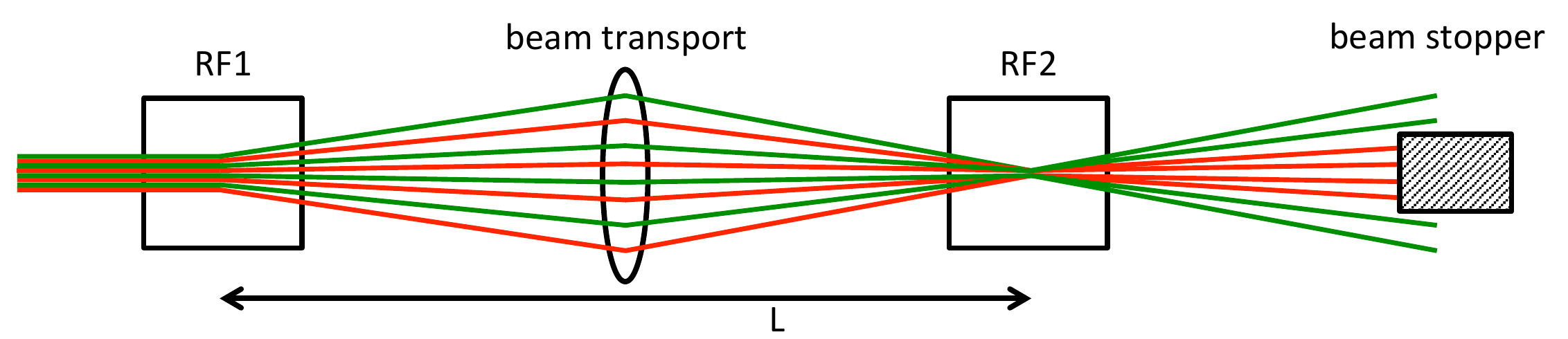}
    \caption{Panofsky-Schnell method for RF-separated beams, exploiting different velocities of particles in a monochromatic beam due to their different masses. The phase between the first (RF1) and second cavity (RF2) is selected in a way to have no deflection in RF2 of unwanted particles (red), which are in turn stopped by an absorber in the beamline. The wanted particles (green) receive a net deflection and pass towards the experiment.}
    \label{fig:Rfbeams}
    \vspace{-1em}
 \end{figure}

The AMBER/COMPASS++ programme comprises a second phase of experiments for strange meson spectroscopy, kaon PDFs and nucleon TMDs via Drell-Yan reactions. All measurements require highest intensities of the wanted secondary antiprotons and kaons that take a minority in the beam composition at the required beam momenta. As the total beam intensity in EHN2 is limited by radiation protection (RP) considerations, the only option to increase the total flux of antiprotons or kaons is to increase their share within the RP limit. One possibility is to exploit velocity differences for different particle species in a monochromatic beam. The Panofsky-Schnell method of separation by radiofrequency (RF) dipole cavities is depicted in Fig.~\ref{fig:Rfbeams} and has been used at CERN several decades ago resulting in low intensity beams. However, recent developments in superconducting RF technology might allow for the development of a high-intensity, high-energy RF-separated beam. Studies for this project are on-going and more information can be found in Ref.~\cite{Gatignon:2650989}.

\vspace{-0.5em}
\section{Conclusions}
\vspace{-0.5em}
The MuonE, AMBER/COMPASS++, and NA64$\mu$ collaborations have proposed several measurements with a very diverse program at the M2 beamline in the North Area of CERN in the framework of the ``Physics Beyond Collider (PBC)'' initiative. Beamline design, feasibility and integration studies have been performed by the EHN2 working group within the PBC Conventional Beams WG. A full report has been published\cite{Gatignon:2650989} together with an executive summary\cite{Gatignon:2650193}, while detailed studies for technical details continue in collaboration with the three proponents. The proposals and letters of intent are currently being discussed and evaluated by the CERN SPSC committee. First test runs can start as early as 2021, when the long stop 2 of CERN's accelerator complex will finish. 

\vspace{-0.5em}
\begin{acknowledgments}
\vspace{-0.5em}
We are thankful for the important input from technical experts at CERN, in particular N. Doble, C.Ahdida, Y.Gaillard, S.Girod, D.Jaillet, V.de Jesus,
J.Lehtinen, and Ph.Schwarz. The studies included in these proceedings have been performed in close collaboration with the representatives of the proposals submitted to PBC and SPSC, in particular V.Andrieux, P.Crivelli, O.Denisov, S.Gninenko, G.Mallot, C.Matteuzzi, and G. Venanzoni. Finally, we acknowledge the excellent collaboration with and support from the PBC management and the chairs of the physics working groups.
\end{acknowledgments}
\vspace{-0.5em}
\vspace{-0.5em}
\bibliography{bibliography}

\end{document}